\documentclass[referee,a4paper,12pt,structabstract]{swsc} 

\usepackage{graphicx}
\usepackage{txfonts}
\usepackage{subfigure}
\usepackage{epstopdf}
\usepackage[backref]{hyperref}
\usepackage{natbib}

\hypersetup{colorlinks=true,citecolor=cyan,urlcolor=cyan,linkcolor=blue}

\begin{document}
%

   \title{Photometric magnetic-activity metrics tested with the Sun: Application to \emph{Kepler} M dwarfs}


   \author{Savita Mathur
          \inst{1}
          \and
          David Salabert\inst{2}
          \and
          Rafael~A. Garc\'ia\inst{2}
          \and
          Tugdual Ceillier\inst{2}
          }

   \institute{Space Science Institute,
              4750 Walnut Street, Suite \#205, Boulder, CO, USA\\
              \email{\href{mailto:smathur@spacescience.org}{smathur@spacescience.org}}
         \and
             Laboratoire AIM, CEA/DSM -- CNRS - Univ. Paris Diderot -- IRFU/SAp, Centre de Saclay, 91191 Gif-sur-Yvette Cedex, France\\
             \email{\href{mailto:david.salabert@cea.fr}{david.salabert@cea.fr};\href{mailto:rgarcia@cea.fr}{rgarcia@cea.fr};\href{mailto:tugdual.ceillier@cea.fr}{tugdual.ceillier@cea.fr}}
             }


 
  \abstract
   {The \emph{Kepler} mission has been providing high-quality photometric data leading to many breakthroughs in the exoplanet search and in stellar physics. Stellar magnetic activity results from the interaction between rotation, convection, and magnetic field. Constraining these processes is important if we want to better understand stellar magnetic activity. 
   }
   {Using the Sun, we want to test a magnetic activity index based on the analysis of the photometric response and then apply it to a sample of M dwarfs observed by \emph{Kepler}.
   }        
   {We estimate a global stellar magnetic activity index by measuring the standard deviation of the whole time series, $S_{\rm ph}$. Because stellar variability can be related to convection, pulsations, or magnetism, we need to ensure that this index mostly takes into account magnetic effects. We define another stellar magnetic activity index as the average of the standard deviation of shorter subseries which lengths are determined by the rotation period of the star. This way we can ensure that the measured photometric variability is related to starspots crossing the visible stellar disk. This new index combined with a time-frequency analysis based on the Morlet wavelets allows us to determine the existence of magnetic activity cycles. 
   }	
   {We measure magnetic indexes for the Sun and for 34 M dwarfs observed by \emph{Kepler}. As expected, we obtain that the sample of M dwarfs studied in this work is much more active than the Sun. Moreover, we find a small correlation between the rotation period and the magnetic index. Finally, by combining a time-frequency analysis with phase diagrams, we discover the presence of long-lived features suggesting the existence of active longitudes on the surface of these stars.
  }	
 {}        

\authorrunning{Mathur et al. 2014}
\titlerunning{Photometric magnetic-activity metrics}
   \maketitle


\section{Introduction}


\noindent The {\it Kepler} mission \citep{2010Sci...327..977B}, which was designed to search for Earth-like exoplanets, monitored 196,468 stars  since its launch in March 2009 \citep{2014ApJS..211....2H}. The mission has already been successful by detecting more than 900 confirmed planets with a variety of parameters and configurations \cite[e.g.][]{2012ApJ...746..123H,2013Natur.494..452B,2014arXiv1402.6534R} and several thousands of planet candidates that await confirmation from ground-based follow up. In addition to the planet investigation, a second scientific  program was part of the mission to study and characterise the stars with asteroseismology. 

\noindent Seismology is the only tool that allows us to directly probe the internal layers of the Sun and the stars. With the excellent quality of the {\it Kepler} data, asteroseismic studies were successfully applied to a large amount of solar-like \cite[e.g.][]{2011A&A...534A...6C,2011Sci...332..213C,2011ApJ...733...95M,2012ApJ...749..152M} and red-giant stars \cite[e.g.][]{2011ApJ...743..143H,2011ApJ...741..119M,2012A&A...537A..30M}. Thanks to the detection of mixed modes in subgiants and red giants \citep{2011Sci...332..205B}, the rotation of their cores could be measured in detail by \citet{2012ApJ...756...19D,2014arXiv1401.3096D}, \citet{2012Natur.481...55B}, and in an ensemble way by \citet{2012A&A...540A.143M}. One major breakthrough in stellar physics was made by \citet{2011Natur.471..608B} who showed that asteroseismology allows us to distinguish between two different evolutionary stages in the red-giant evolution: giants that are burning hydrogen in a shell and those that started burning helium in the core.

\noindent Seismology can also provide information on stellar magnetic activity. As it is well known for the Sun, magnetic activity impacts the p-mode characteristics: indeed as the magnetic activity increases, the mode frequencies increase while their amplitudes decrease. 
\citet{2010Sci...329.1032G} observed the same behaviour in the CoRoT \citep[Convection, Rotation, and planetary Transits;][]{2006cosp...36.3749B} data and detected for the first time magnetic activity in a solar-like star with asteroseismic analyses. Further analyses showed that similarly to the Sun, the shift of the frequencies is larger for high-frequency modes \citep{2011A&A...530A.127S}, suggesting that the change in activity happens in the outer layers of the star rather than in its deeper layers.



\noindent So far, the detailed mechanisms generating the solar magnetic activity are not completely understood. For instance, the long minimum between cycles 23 and 24 was not predicted by dynamo models \citep{2006ApJ...649..498D}.  Stellar magnetic activity results from the interaction between differential rotation, convection and magnetic field \citep[e.g.][]{2011ApJ...742...79B}. One way to better grasp the details of the processes in play is to study other stars with different conditions (different ages, rotation periods, masses...) in order to have a broader vision \citep[e.g.][]{2011arXiv1111.2065M}.  Seismology is a powerful tool that can not only detect magnetic activity but also provide the internal structure and dynamics of stars (depth of the convection zone, rotation profile, convection properties), which are all crucial components in the understanding of the solar and stellar magnetic activity.  We also note that the {\it Kepler} data allow us to measure differential rotation as shown recently by \citet{2013A&A...560A...4R}.


\noindent Large spectroscopic surveys contributed to the study of stellar photospheric and chromospheric activity because magnetic fields affect the electromagnetic spectrum lines (e.g. CaII, H$\alpha$, Na). Surveys such as the Mount Wilson HK project \citep{1978ApJ...226..379W} or the one led at the Solar-Stellar Spectrograph \citep{2007AJ....133.2206H} collected several decades of time series for hundreds M- to F-type stars for  \citep[see][for a review]{2008LRSP....5....2H}. These observations showed that there is a large variety of behaviours in terms of magnetic activity \citep[e.g.][]{1995ApJ...438..269B,1998ApJS..118..239R} with stars showing regular or irregular cycles or a quite constant magnetic regime. Using the Mount Wilson observations, \citet{2002AN....323..357S} highlighted the relationship between the magnetic cycle period and the rotation period. Hence, fast rotators have shorter magnetic cycles. To understand this variety of behaviours we need to relate them with differences in stellar parameters (mostly with their internal structure and composition). For instance, hot F-type stars seem to have more irregular cycles than G-type stars. Spectropolarimetric observations provide additional constraints by measuring the stellar magnetic field and its topology. For instance, magnetic polarity reversal was observed for tau Boo by \citet{2009MNRAS.398.1383F}. Recently a large spectropolarimetric survey of FGK type stars was led by the Bcool team \citep{2013arXiv1311.3374M} with the measurement of the magnetic field in $\sim$\,70 solar-type stars. They found that cool stars have stronger magnetic field than hot ones. Finally, detection of magnetic fields in red giants was also done in several stars \citep[e.g.][]{2012A&A...541A..44K}. 


\noindent Photometric observations can also provide information on the magnetic activity of the stars as done for instance by \citet{2012ARep...56..716S} with M dwarfs. In the past, several teams performed time-frequency analyses of photometric or spectroscopic datasets either with wavelets \citep{1997ApJ...483..426F} or with short-term Fourier transforms \citep{2009A&A...501..703O}, detecting a variety of cycles evolving with time. Recently, different indexes have been defined by \citet{2010ApJ...713L.155B} to characterise the variability of {\it Kepler} targets using photometric observations. They defined the ``range'', $R_{\rm var}$, and considered it in \citet{2013ApJ...769...37B} as a metric of the photometric magnetic activity. However as this metric takes into account values of the stellar flux between 5\% and 95\% of the brightness span, it can underestimate the activity level of very active stars. Another index defined by \citet{2013ApJ...769...37B} is the median differential variability, MDV, computed on data that are rebinned from 1 hour up to 8 days. But we have to be careful that the variability in the light curves can be due to different phenomena such as pulsations, granulation, rotation, or the presence of spots. To specifically study variability induced by magnetic activity and define an indicator that properly measures it, we need to take into account the rotation period of the star as its measurement relies on the presence of spots (thus magnetic field) on the stellar surface. In addition, these indexes are computed on subseries of 30 days, which can bias the results for stars rotating much slower than 15 days. This is the reason why we define and use a different metric in this work.

\noindent In Section 2, we describe the solar data and the sample of M dwarfs observed by {\it Kepler} that are used in this work. These M dwarfs are a subsample of the stars presented in \cite{2013MNRAS.432.1203M}. They are fully convective very low-mass stars (0.3 to 0.55$M_{\odot}$), with an effective temperature $T_{\rm eff}$ smaller than 4000K and a surface gravity $\log g \ge 4$. Even though we haven't detected oscillation modes in these stars, they represent a good benchmark for testing our magnetic indexes due to their high-activity levels. Furthermore, as said above, magnetic activity cycles have already been detected in M dwarfs in the past.
We define in Section 3 a magnetic activity index based on the analysis of the light curves and the prior knowledge of the stellar rotation periods. We measure this index in the case of the Sun and a few tens of M dwarfs. Section 4 discusses the magnetic activity of M dwarfs. Finally Section 5 presents our conclusions.

\section{Observations and Data Analysis}

In this work we use data collected by two space missions. On one hand, $\sim$ 3.5 years (quarters 1 to 15) of continuous observations of the NASA {\it Kepler} mission allowed us to measure the long-term variability of a sample of stars in the constellation of Cygnus and Lyra. At any time, {\it Kepler} observed around 120,000 stars, most of them at a cadence of 29.42 minutes \citep[e.g.][]{2010ApJ...713L.160G}. Due to the orbital configuration, and to maintain the solar panel properly oriented towards the Sun, {\it Kepler} experiences a roll every 90 days (a quarter of a year). Data were consequently subdivided into quarters and the observed stars were placed in a different CCD in the focal plane. To downlink these data,  {\it Kepler} pointed to the Earth in a monthly basis, which introduces gaps in the recorded time series, as well as some thermal drifts and other instrumental instabilities \citep[e.g.][]{2010ApJ...713L.120J}. In this work we used NASAÕs Simple Aperture Photometry (SAP) light curves \citep{2010ApJ...713L.120J} that have been corrected of outliers, jumps, and drifts following the methods described in \cite{2011MNRAS.414L...6G}. Based on the {\it Kepler} Input Catalog \citep{2011AJ....142..112B}, the M dwarfs selected for this work have magnitudes between 13 and 16, $T_{\rm eff}$ between 3600 and 4000\,K, and $\log g$ between 4 and 4.6\,dex. 

\noindent On the other hand, we use 16 years of continuous photometric observations of the Sun recorded by the Variability of solar IRradiance and Gravity Oscillations \citep[VIRGO,][]{1995SoPh..162..101F} instrument aboard the ESA-NASA Solar and Heliospheric Observatory \citep[SoHO,][]{DomFle1995}. VIRGO is composed of several instruments. In particular, the sum of the green and red channels of the Sun PhotoMeters (SPM) are a good photometric approximation of the {\it Kepler} bandwidth \citep{2010ApJ...713L.155B}. It can be used to study the magnetic activity properties of the Sun and as a reference for other stars observed by {\it Kepler}. The standard VIRGO/SPM data are high-pass filtered with a cut-off frequency of a few days. In order to monitor the solar variability at the frequencies of the rotation, we used raw VIRGO/SPM data processed with the procedures described in \cite{GarSTC2005,2011MNRAS.414L...6G} in a similar way to what we do with {\it Kepler}.


\section{Photometric Index of Magnetic Activity}


To measure the magnetic activity of a star through the variability observed in its light curve, we can first use a global index as defined by \cite{2010Sci...329.1032G}, which is simply the standard deviation of the whole light curve. Hereafter we call this index $S_{\rm ph}$. This is based on the presence of starspots on the stellar surface.

\begin{figure}
   \centering
   \includegraphics[width=0.9\textwidth]{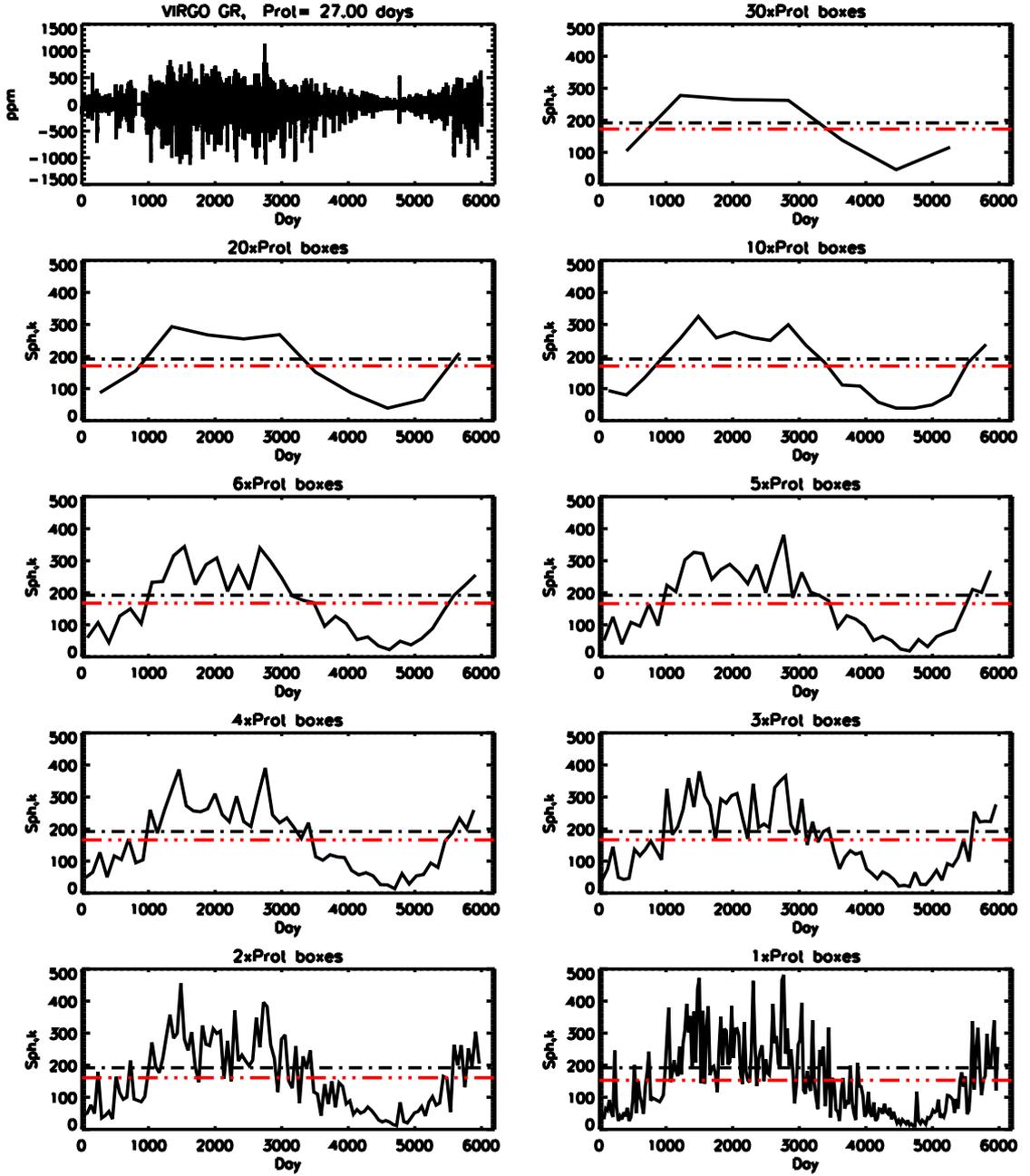}
      \caption{Time series from the VIRGO/SPM instrument obtained with the green and red channels as described in Section 3.1 (top left panel). Standard deviation for VIRGO data using subseries of size $k \times P_{\rm rot}$ with $k$=1, 2, 3, 4, 5, 6, 10, 20, and 30 (from bottom right to top right panel). The black dot dash line represents the global magnetic index $S_{\rm ph}$ and the red triple dot dash line corresponds to the mean magnetic index, $<S_{\rm ph, k}>$, computed as described in section 3.2.
              }
         \label{FigVIRGO}
   \end{figure}

\noindent As said above, the surface magnetic activity is related to an inner dynamo process linked to the rotational period of the star. Thus, when defining an index to study stellar magnetic activity on a large sample of stars, as provided by {\it Kepler}, the stellar rotation is a key parameter,  if the rotation period is accurately estimated \citep{2013JPhCS.440a2020G,2013MNRAS.432.1203M}. It is also natural to define an index which takes into account possible temporal variations of the activity level. To do this, a given light curve is divided into subseries of $k \times P_{\rm rot}$, where $P_{\rm rot}$ is the rotational period of the star. Values of $k$ between 1 and 30 were tested and the analysis was performed on the Sun and some active stars. For each individual subseries, the standard deviation $S_{\rm ph,k}$ of the non-zero values (i.e. points that are not missing) is calculated, providing the possibility to study any temporal evolution of the star's activity level (see Figure~\ref{FigVIRGO}).  The mean value of the standard deviations $<S_{\rm ph, k}>$ represents a global magnetic activity index, comparable to $S_{\rm ph}$, except that it takes into account the rotation of the star. This method also allows us to measure the magnetic index during the minimum or maximum period of the magnetic cycle by only taking the mean of the subseries that have a standard deviation smaller (or larger) than the global index.
Finally the returned value is corrected from the photon noise. In the case of the Sun, we use the high-frequency part of the power spectrum to estimate it, while the magnitude correction from \cite{2010ApJ...713L.120J} is used to estimate the corrections to apply to the {\it Kepler} stars.

\subsection{Sun}
Figure~\ref{FigVIRGO} illustrates this calculation in the case of the Sun using $\sim$~6000 days of the photometric VIRGO/SPM observations (top-left panel of Figure~\ref{FigVIRGO}), which were rebinned into a 30-min temporal sampling in order to mimic the long-cadence {\it Kepler} data. The temporal variation of the magnetic index $S_{\rm ph,k}$ was calculated for different values of the factor $k$: [30, 20, 10, 6, 5, 4, 3, 2, and 1] $\times  P_{\rm rot}$, with a solar $P_{\rm rot}$ taken to be equal to 27.0 days (individual panels on Figure~\ref{FigVIRGO}). The black dot-dashed line represents the value of the standard deviation over the entire time series, $S_{\rm ph}$, and the red triple dot-dashed line the mean value of the standard deviations, $<S_{\rm ph, k}>$, calculated for each $k$. The well-known 11-yr solar cycle is well reproduced with such activity index, allowing us to track shorter time scaled features, like for instance the double peak observed during solar maxima (between day $\sim$~1000 and day $\sim$~ 3000 on Figure~\ref{FigVIRGO}). Note that the value of $<S_{\rm ph, k}>$ is slightly smaller than $S_{\rm ph}$, and it converges towards $S_{\rm ph}$ as the value of $k$ is increased.


 \begin{figure}
   \centering
   \includegraphics[width=0.9\textwidth]{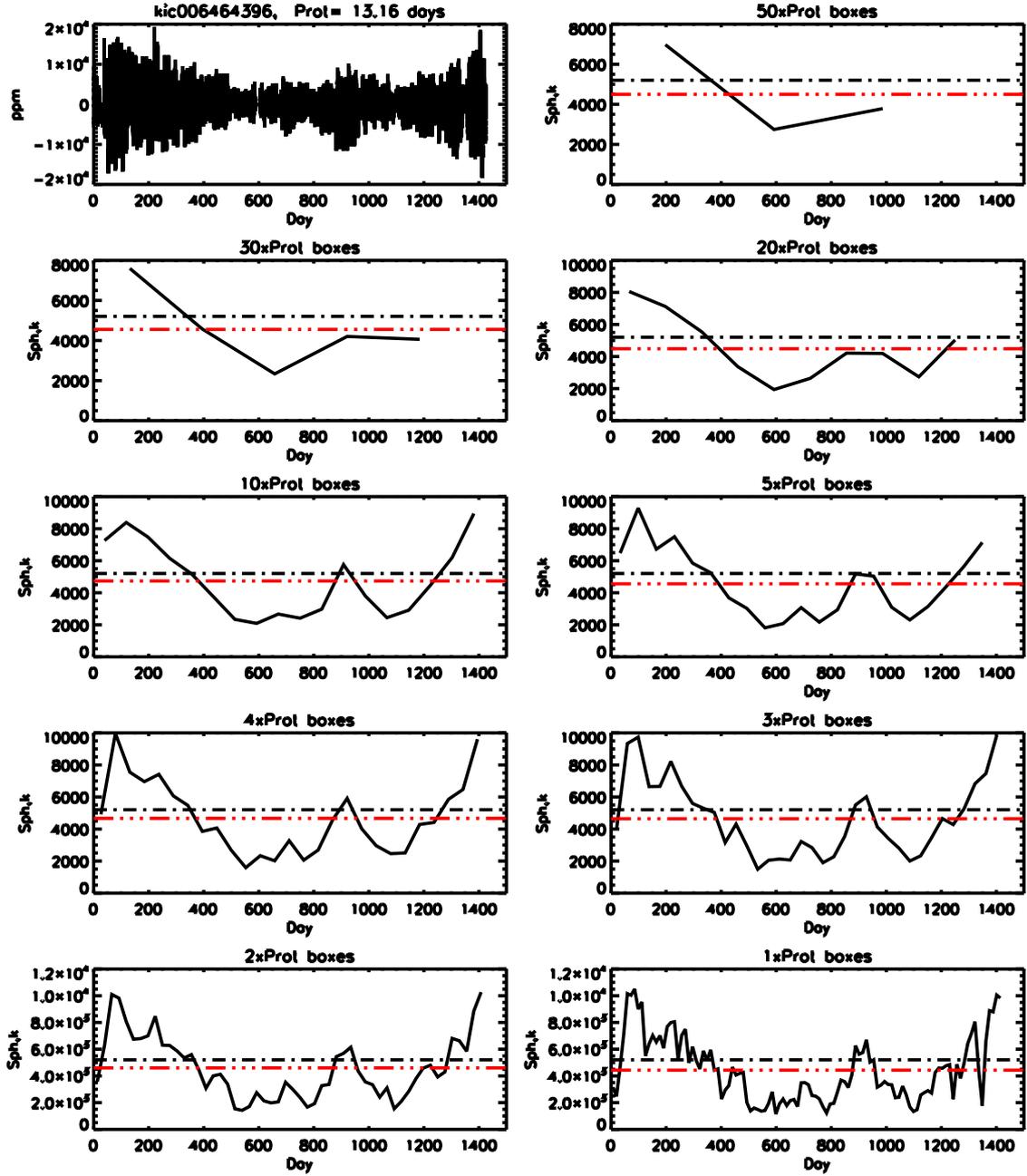}
      \caption{Temporal variation of the magnetic index $S_{\rm ph, k}$ for the M dwarf KIC~6464396. Same legend as in Figure~\ref{FigVIRGO}.
              }
         \label{FigM}
   \end{figure}

\subsection{M dwarfs observed by Kepler}

M dwarfs are know to be very active stars presenting a large number of flares. Thus they provide a very good benchmark to test the magnetic indexes described above.
In order to determine the optimal value of the factor $k$, we apply the same analysis to a sample of 34 magnetically active M stars observed by {\it Kepler} for which $\sim$1300 days (Q1-Q15)  of continuous observations are available. These stars were chosen to have a rotation period shorter than 15 days as measured by \cite{2013MNRAS.432.1203M}. They obtained the rotation periods by applying the autocorrelation function on 10 months of data. We checked their values with a time-frequency analysis based on wavelets \citep{1998BAMS...79...61T,2010A&A...511A..46M} that we performed on the whole timeseries. The rotation periods for the 34 stars agree between the two methods within the error bars of the wavelets.  Since these stars are rather faint, there is a possibility that the light curves can be polluted by a nearby companion. We checked the crowding and the pixel data of all the stars. The crowding values are listed in Table 1 and are above 0.8 for most of the stars except three. The inspection of the pixel data suggests that these three stars (with a ``NO'' flag in Table 1) are most likely polluted by a nearby star.

\begin{figure}[h!]
   \centering
   \includegraphics[width=8cm]{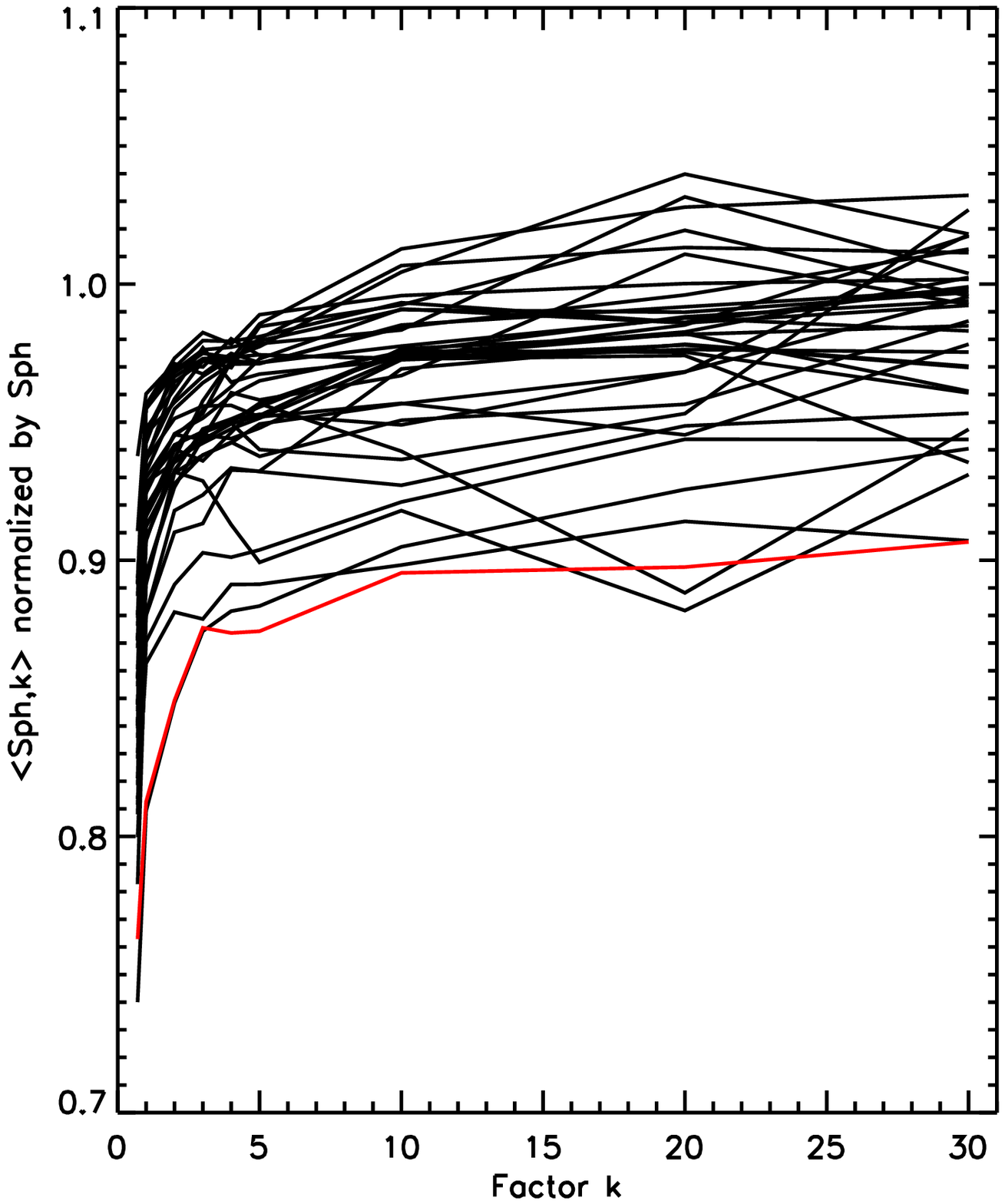}
   	\includegraphics[width=8cm]{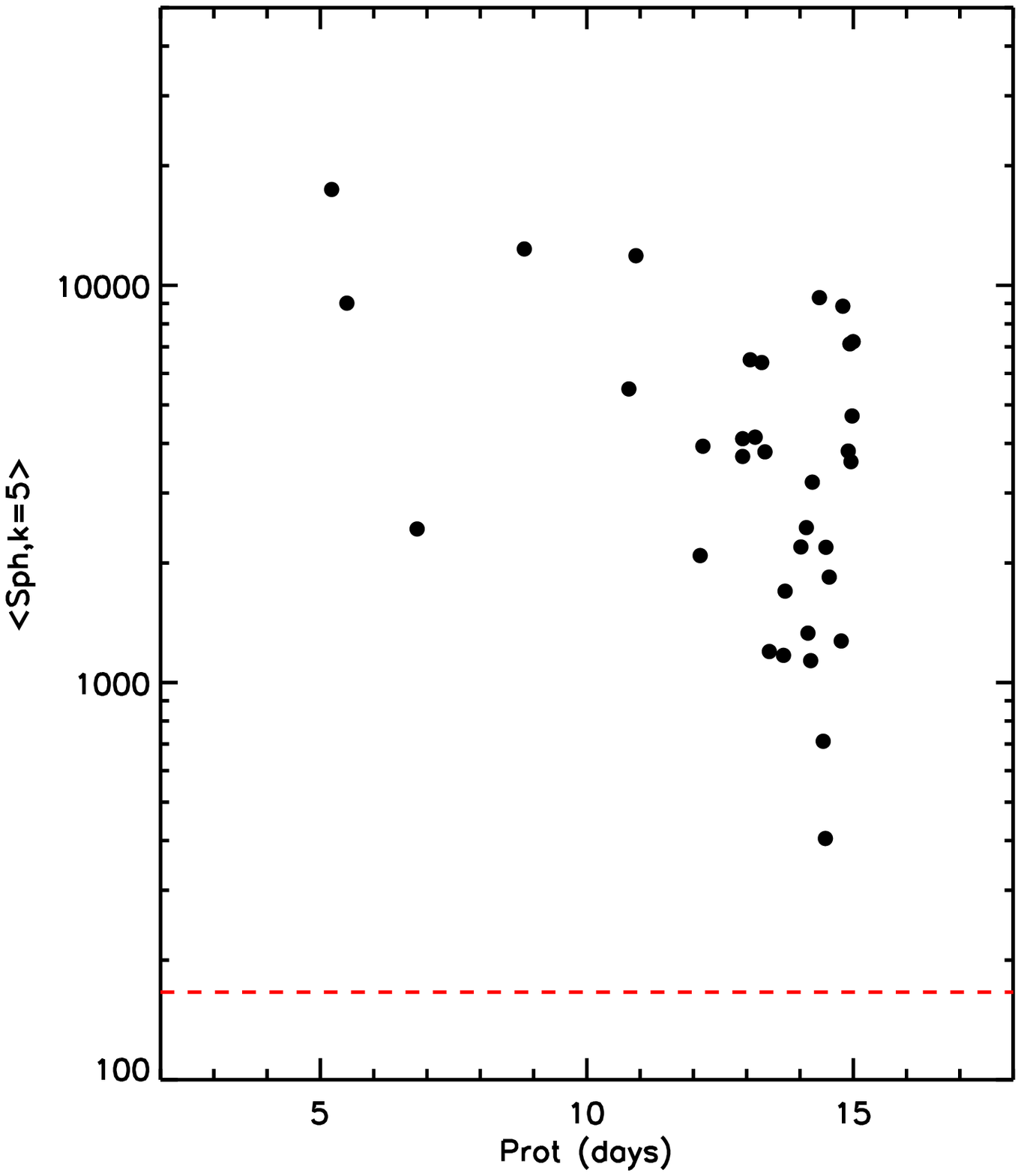}
      \caption{Left panel: Normalised mean standard deviation as a function of $k$, which is the multiplicative factor of $P_{\rm rot}$ to determine the size of the subseries, for the 34 M dwarfs of our sample (black curves) and the Sun (red curve). Right panel:  Magnetic activity index for the value $k$=5, $<S_{\rm ph, 5}>$, corrected from the photon noise, as a function of the rotation period for the 31 non-polluted M dwarfs. The red dashed line corresponds to the solar value.
              }
         \label{Fig2}
   \end{figure}

\noindent The indices $<S_{\rm ph, k}>$ are given in Table~\ref{default}. The calculation for an example star, KIC~464396, is shown in Figure~\ref{FigM} for different values of $k$. We also corrected the indexes for the photon noise by following the relation established by \cite{2010ApJ...713L.120J}. The left panel of Figure~\ref{Fig2} shows $<S_{\rm ph, k}>$  normalized by the standard deviation of the whole time series $S_{\rm ph}$ as function of the factor $k$ for the 31 M stars. The values in the case of the Sun are represented in red. The ratio $<S_{\rm ph, k}>$ / $S_{\rm ph}$ tends to become constant and close to 1 for higher $k$. A value of 5 $\times P_{\rm rot}$ appears to reasonably describe the magnetic temporal evolution of stars as well as to give a correct value of global activity index. 

\noindent The right panel of Figure~\ref{Fig2} shows the mean activity indexes for the 31 non-polluted M stars in the case of 5 $\times P_{\rm rot}$, $<S_{\rm ph, k=5}>$, as a function of $P_{\rm rot}$. It is clear that the M stars present a wide range of magnetic activity as in the selected sample the most active star is about 20 times more active than the less active star, while $<S_{\rm ph, k=5}>$ is 4 to 80 times greater than the solar value of 166.1\,$\pm$\,2.6\,ppm \citep{2014A&A...562A.124M}. 
\citet{1984ApJ...279..763N} showed that there is a correlation between the Ca II HK flux, $R'_{HK}$ index, and the rotation period. We obtain a slightly similar trend where fast rotators have larger magnetic indexes.

\begin{table}[htdp]
\caption{List of M dwarfs analysed with the rotation periods from \citet{2013MNRAS.432.1203M}, the magnetic indexes $<S_{\rm ph, k}>$ for $k$=5 and $S_{\rm ph}$ corrected from the magnitude following \citet{2010ApJ...713L.120J}, the crowding, and a flag for possible pollution from a nearby star.}
\begin{center}
\begin{tabular}{ccccccc}
\hline
Ref. \# & KIC & $P_{\rm rot}$ (days) & $<S_{\rm ph, k=5}>$ (ppm) & $S_{\rm ph}$ (ppm)  & Crowding & Flag\\
\hline
\hline
 1&   2157356 &  12.9&  4109.1&  4445.1& 0.85& OK\\
 2&   2302851 &  12.2&  3934.1&  4148.9& 0.96& OK\\
 3&   2570846 &  10.9& 11871.9& 12481.5& 0.88& OK\\
 4&   2574427 &  13.4&  1196.9&  1245.5& 0.79& NO\\
 5&   2692704 &  14.8&  8858.5&  9183.6& 0.85& OK\\
 6&   2832398 &  15.0&  4688.7&  4761.5& 0.82& OK\\
 7&   2834612 &  13.3&  6391.3&  6591.1& 0.89& OK\\
 8&   2835393 &  15.0&  3597.7&  3705.4& 0.85& OK\\
 9&   3102763 &  14.4&  9310.4&  9960.1& 0.89& OK\\
10&   3232393 &  14.5&   404.9&   427.0& 0.90& OK\\
11&   3634308 &  12.9&  3708.0&  3895.9& 0.88& OK\\
12&   3935499 &   5.2& 17429.4& 19318.5& 0.94& OK\\
13&   4833367 &  14.2&  1135.2&  1171.6& 0.90& OK\\
14&   5041192 &  10.8&  5483.9&  5549.3& 0.90& OK\\
15&   5096204 &  14.8&  1272.1&  1315.2& 0.92& OK\\
16&   5210507 &   8.8& 12345.3& 12899.8& 0.87& OK\\
17&   5611092 &  14.4&   711.4&   738.7& 0.94& OK\\
18&   5900600 &  14.0&  2194.3&  2307.9& 0.92& OK\\
19&   5950024 &  14.1&  2454.5&  2587.0& 0.96& OK\\
20&   5954552 &  14.9&  7121.1&  7314.2& 0.95& OK\\
21&   5956957 &  14.9&  3825.4&  4027.0& 0.94& OK\\
22&   6307686 &  13.3&  3807.7&  4029.1& 0.94& OK\\
23&   6464396 &  13.2&  4146.5&  4694.7& 0.90& OK\\
24&   6545415 &   5.5&  9018.6& 10233.5& 0.57& NO\\
25&   6600771 &  13.1&  6494.6&  6845.9& 0.92& OK\\
26&   7091787 &  14.1&  1331.6&  1523.7& 0.92& OK\\
27&   7106306 &  14.2&  3194.6&  3260.4& 0.95& OK\\
28&   7174385 &  14.5&  2190.2&  2229.1& 0.94& OK\\
29&   7190459 &   6.8&  2435.6&  2631.1& 0.95& OK\\
30&   7282705 &  14.5&  1842.5&  1891.7& 0.83& OK\\
31&   7285617 &  13.7&  1698.3&  1746.5& 0.69& NO\\
32&   7534455 &  12.1&  2087.3&  2198.0& 0.94& OK\\
33&   7620399 &  13.7&  1170.5&  1202.1& 0.92& OK\\
34&   7673428 &  15.0&  7213.6&  7703.9& 0.91& OK\\

\hline
\end{tabular}
\end{center}
\label{default}
\end{table}%




\section{Magnetic activity of M dwarfs}

\begin{figure}
   \centering
   \includegraphics[width=12cm, trim=0 0 4cm 1.5cm]{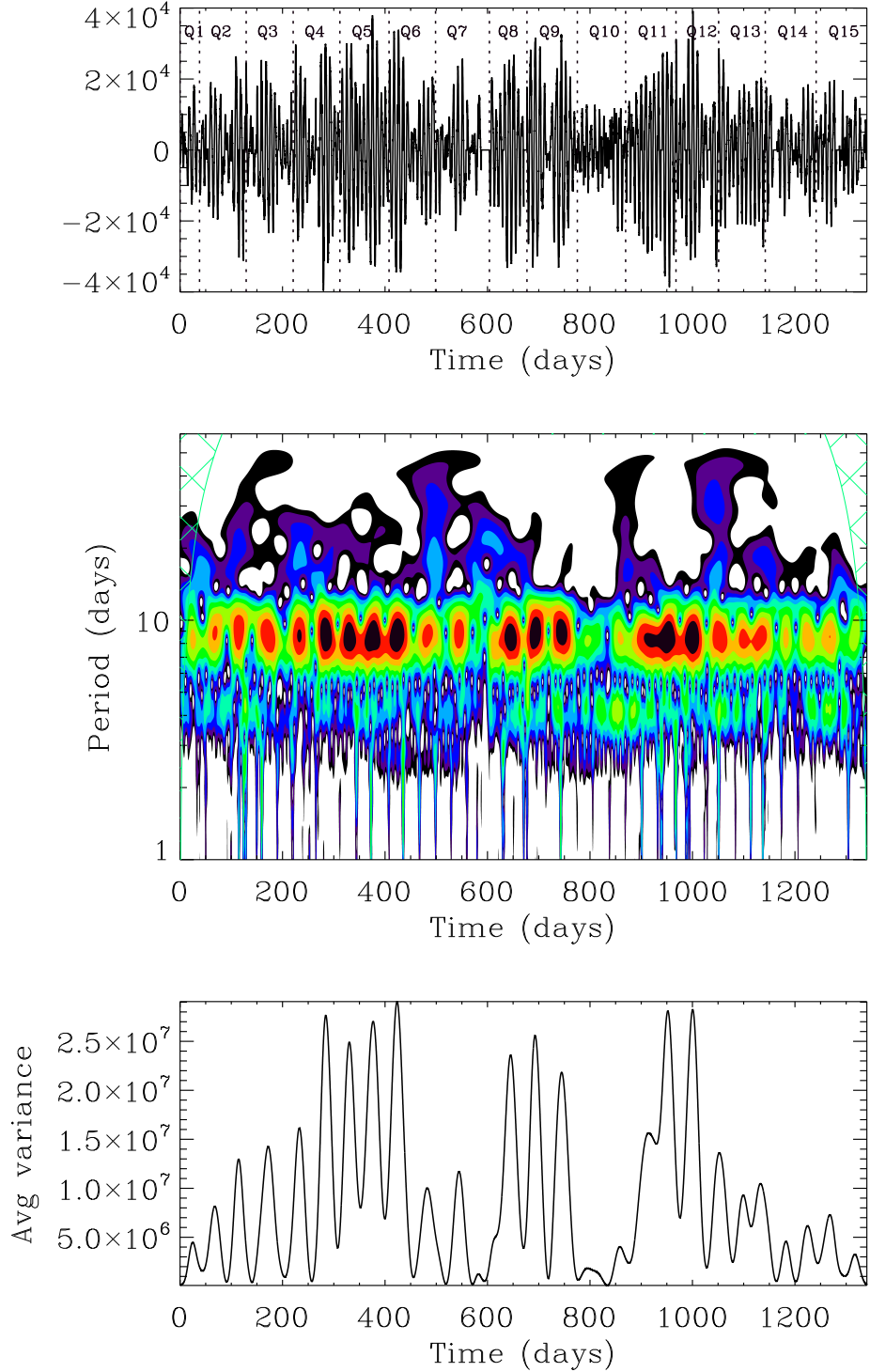}
      \caption{Top panel: time series of the star KIC~5210507 corrected as described in Section 2. Middle panel: Wavelet Power Spectrum as a function of time and period. Dark and red colours correspond to high power while blue and purple colours correspond to low power. The green grid corresponds to the cone of influence that delimits the reliable regions of the WPS by taking into account edge effects. Bottom panel: Scale-average variance obtained by projecting the WPS on the time axis around the rotation period of the star (8.43 days).
              }
         \label{Fig3}
   \end{figure}

When performing the time-frequency analysis, we need to distinguish between different phenomena: pulsation modes, a small differential rotation leading to a beating that mimics a magnetic activity cycle, and a real magnetic activity cycle. As pointed out by \cite{2013MNRAS.432.1203M}, the auto-correlation function that they use to determine the rotation periods can detect acoustic modes revealing a few red giants in their sample. In all the stars of our sample, the time-frequency analysis based on the wavelets shows a modulation that could be attributed to a magnetic activity cycle. This analysis consists of measuring the correlation between the Morlet wavelet (the product of a sinus wave and a Gaussian function) and the time series by sliding the wavelet along it and by changing the period of the wavelet in a given range.  The wavelet power spectrum (WPS) obtained is shown in the middle panel of Figure~\ref{Fig3} for the M dwarf KIC~5210507. We detect a rotation period of 8.43\,$\pm$\,0.69\,days. The inspection of the pixel data confirms that there is no pollution from a close companion. The detailed analysis of the power spectrum revealed the presence of several peaks around the inferred rotation period suggesting the existence of latitudinal differential rotation on the surface of this star. We retrieve a relative differential rotation of 20\%, which is lower than the value of 30\% for the solar case. This is slightly larger than what would be expected theoretically following the work of \cite{2011AN....332..933K}. Most of the other stars in our sample show a complicate peak structure around the rotation period that could also be indicative of the presence of surface differential rotation. Unfortunately, in the wavelet power spectrum we do not have enough resolution to measure it. The detailed analysis of the differential rotation is out of the scope of this paper and will be the object of future investigation.

\begin{figure}
   \centering
   \includegraphics[width=10cm, angle=90]{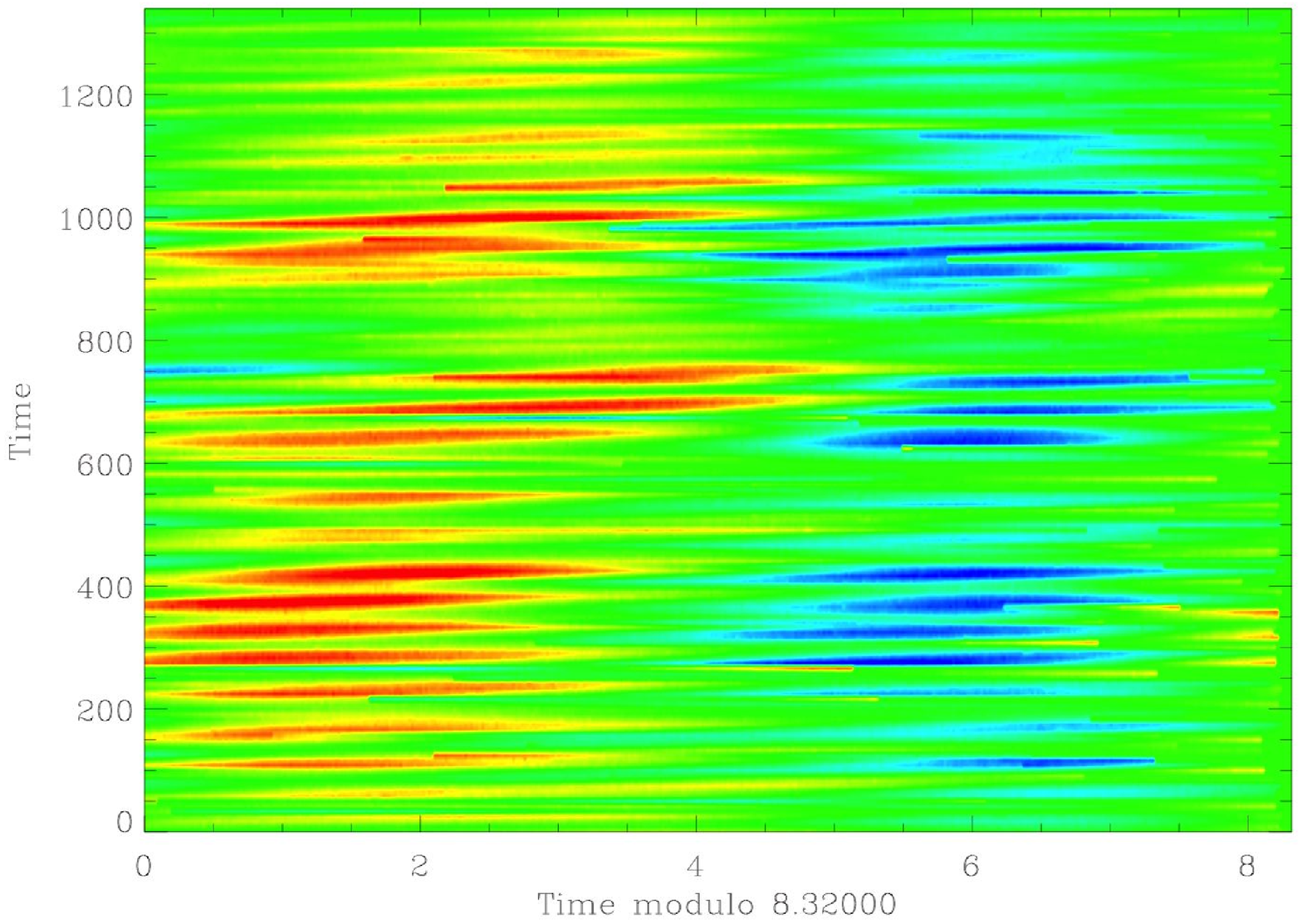}
      \caption{Phase diagram as described in Section 4 with the time modulo a value close to the rotation period of the star on the x-axis and the time on the y-axis.
              }
         \label{Fig4}
   \end{figure}

\noindent The lower panel of Figure~\ref{Fig3} represents the scale-average variance that corresponds to the projection of the WPS on the time axis. This provides a good way to study the temporal evolution of the magnetic activity as shown by \cite{2013JPhCS.440a2020G}. We notice a modulation in the main activity level of $\sim$45 days. Exploring in more details the peak structure around the rotation period of KIC~5210507 we found two main peaks at 9.92 days (1.167$\mu$Hz) and  8.33 days (1.390\,$\mu$Hz). The beating between these two periods can produce a modulation of 50 days, very close to what is observed in the light curve. In order to have an efficient beating between the two frequencies, their phases need to be nearly constant during a long period of time. As long as a phenomenon -- spot, active
longitude or pulsation -- lives  and its frequency does not evolve with time, the phase of the modulation must
remain the same. We thus follow a similar method to the one used to
study TTV \citep[Transit Timing Variations,][]{2011ApJ...727...24T}: we cut the light curve into bits
of length equal to the period of the modulation and we stack these bits one on
top of the other, in an \'echelle-like diagram \citep{1983SoPh...82...55G}. The amplitude of each
data point is translated into a color magnitude. By doing so, we
obtain a figure showing the phase of each occurrence of the modulation
through the whole light curve. Every vertical ridge in this figure
corresponds to a stable phase, the lifetime of the phenomenon being
the vertical extension of the ridge. An example of the phase diagram for KIC~5210507 is shown in Figure~\ref{Fig4}. We clearly see that the maxima (in red) and the minima (in blue) produce vertical ridges, thus a stable phase during more than 1000 days. In the case of spot migration with a surface differential rotation, we would expect to observe a gradual shift of the phase. What we observe could suggest that there are active longitudes where the spots have a preferred longitude to emerge at the surface of the star \citep[e.g.][]{2006A&A...445..703B,2013ApJ...770..149W}.

\section{Conclusion}

We have defined two magnetic indexes that are measured on photometric observations. The first one, $S_{\rm ph}$, is calculated as the standard deviation of the full-length time series. It provides a mean activity level during the observation period. The second one, $<S_{\rm ph, k}>$, is a magnetic index based on the knowledge of the stelar surface rotation period, $P_{\rm rot}$, by smoothing the time series over $k \times P_{\rm rot}$. We computed $<S_{\rm ph, k}>$ for different values of $k$ and we tested it on solar data collected during $\sim$6000 days by  the VIRGO/SPM instrument aboard SoHO. Then we applied it to our sample of M dwarfs observed by  \emph{Kepler}. We showed that $k$=5 is a good choice to keep the information on the global magnetic activity while having short enough subseries to track any cycle-like variations in the mean activity level. For larger values of $k$, the index reaches saturation. This analysis shows in particular that M dwarfs are more active than the Sun confirming H$\alpha$ observations at the McDonald Observatory \citep{2013ApJ...764....3R}.

\noindent We found a slight anti-correlation between $<S_{\rm ph, k}>$ and $P_{\rm rot}$ where fast rotators seem to be more active than the slow rotators. This trend agrees with the findings of \citet{1984ApJ...279..763N} based on the observations of CaHK for stars with different spectral type and more recently by \citet{2013arXiv1311.3374M} using a spectropolarimetric survey.

\noindent We also performed a time-frequency analysis for all the stars of our sample. We detected the signature of latitudinal differential rotation suggested by the presence of several peaks around the rotation period. In some stars, we demonstrate that the beating of some of these high amplitude peaks leads to a  modulation in the mean activity level of the light curve that could be misinterpreted as a magnetic activity cycle. However, we could not detect any sign of spots migration.  The computation of the phase diagram shows the existence of long-lived features at their surface and thus of active longitudes which has not been observed in this type of stars. Their existence in M dwarfs and solar-like stars suggest that the outer convective zone plays a role in the mechanism responsible for the development of the surface magnetic features.

\noindent Recently, a similar analysis was applied to F-type solar-like stars \citep[see][]{2014A&A...562A.124M} ($T_{\rm eff} \ge 6000K$ and more massive than the Sun) for which asteroseismic studies could provide key information on their internal structure and dynamics. By combining all this information for a large number of stars --with different structures and dynamics-- it will greatly contribute to a better understanding of the dynamo mechanisms. In addition, by inferring more precise stellar ages (thanks to asteroseismology for instance), we will be able to define a more accurate relationship between age-activity-rotation.

\begin{acknowledgements}
      This work was partially supported by the NASA grant NNX12AE17G. SM, TC and RAG acknowledge the support of the European CommunityÕs Seventh Framework Program (FP7/2007-2013) under grant agreement no. 269194 (IRSES/ASK) and no. 312844 (SPACEINN). RAG acknowledges the support of the French ANR/IDEE grant. DS acknowledges the support provided by CNES.

\end{acknowledgements}

\bibliographystyle{aa} 
\bibliography{/Users/Savita/Documents/BIBLIO_sav}


\end{document}